\documentclass[aps,prl,floatfix,superscriptaddress,twocolumn,footinbib]{revtex4-1}
\usepackage{amssymb}
\usepackage{amsmath}
\usepackage{amsfonts}
\usepackage{appendix}
\usepackage{bm}
\usepackage{graphicx}
\usepackage{epsfig}
\usepackage{epstopdf}
\usepackage{balance}
\usepackage[dvipsnames]{xcolor}
\usepackage{calc}
\usepackage{natbib}
\usepackage[colorlinks,
            linkcolor=blue,
            anchorcolor=blue,
            citecolor=blue,
            urlcolor=blue]{hyperref}
\usepackage{lipsum}

\makeatletter

\newcommand{\Rmnum}[1]{\expandafter\@slowromancap\romannumeral #1@}
\makeatother

\begin{document}

\title{Vibration-Enhanced Spin-Selective Transport of Electrons in DNA Double Helix}
\author{Gui-Fang Du}
\affiliation{School of Physics and Wuhan National High Magnetic Field Center,
Huazhong University of Science and Technology, Wuhan 430074,  China}
\author{Hua-Hua Fu}
 \email{hhfu@mail.hust.edu.cn}
 \affiliation{School of Physics and Wuhan National High Magnetic Field Center,
Huazhong University of Science and Technology, Wuhan 430074,  China}
\author{Ruqian Wu}
\email{wur@uci.edu}
\affiliation{Department of Physics and Astronomy, University of California, Irvine, California 92697-4575, USA}

\begin{abstract}
The spin-selective transport through helical molecules has been a hot topic in condensed matter physics, because it develops a new research direction in spintronics, \emph{i.e.}, chiro-spintronics. Double-stranded DNA (dsDNA) molecules have been considered as promising candidates to study this topic, since the chiral-induced spin selectivity (CISS) effect in dsDNA was observed in experiment. Considering that the dsDNA molecules are usually flexible in mechanical properties, vibration may be one of important factors to influence the CISS effect. Here, we investigate the influences of electron-vibration interaction (EVI) on the spin-selective transport in dsDNA molecules. We uncover that the EVI not only enhances the CISS effect and the spin polarization ($P_s$) in dsDNA, but also induces a series of new spin-splitting transmission modes. More interesting, these vibration-induced transmission spectra tend to host the same $P_s$ values as those of the original spin-splitting transmission modes, making the $P_s$ spectra to display as a continuous platform even in the energy gap. Our work not only provides us a deep understanding into the influence of vibrations on the CISS effect in helical molecules, {but also puts forwards a feasible route to detect the vibration-induced spin-polarized transport in low-dimensional molecular systems}.

\end{abstract}
\maketitle

\section{\Rmnum{1}.  INTRODUCTION}

Recently, molecule spintronics has attracted remarkable research interests since the spin-filtering effect (SFE) and spin-polarization transport were observed in double-stranded DNA (dsDNA) molecules and other helical oligopeptides in experiments \cite{Science_paper,CSR_paper,JPCL_paper,JPCC_paper,JACS_paper,ACS_Nano_paper,Adv_Mater}. The SFE in organic molecules not only provides us a way to manipulate the spin degrees of freedom of electrons, but also inspires the search for a new class of materials to build spintronic devices, as organic molecules usually preserve long spin relaxation time and may self-assemble on different substrates. {Among the possible mechanisms, the helical-induced spin-orbital coupling (SOC) has been identified as the key factor in generating spin polarization in helical molecules \cite{PRL_paper_Sun_1,PRB_paper_Sun_1,PRB_paper_Sun_2,PRB_paper_Gut,JPCL_paper_2,JPCC_Gut_1,JPCC_Geyer_1,PRL_paper_Sun_1,PNAS_Sun_paper}. Thus, the chiral-induced spin selectivity (CISS) has been recognized as an important research frontiers, giving birth to a new research direction in spintronics: chiro-spintronics or chiral-based spintronics \cite{Small_Method}}. It is inspiring that based on the CISS effect, the spin-resolved currents can be generated and controlled at a molecular level, if chiral molecules are utilized as spin-specific transport media.

To develop chiro-spintronic devices, one of the crucial conditions is to enhance CISS effect or to achieve high spin-polarized transport in helical molecules. To this end, several effective ways such as applying an external gate voltage \cite{PRB_paper_Sun_1} and using magnetic helix \cite{PRB_Mag_helix} have been put forwards in theory. These proposed ways are tightly related with unique molecular structures, such as the double helixes in dsNDA molecules. Considering the fact that most helical molecules are usually flexible in mechanical property, lattice disordering and electron dephasing may easily occur in these spintronic devices, leading to the lose of information. This inspires us to explore new physical mechanism that may enhance the CISS effect and spin-polarized transport in dsDNA molecules. Moreover, just due to the aforementioned structural flexibility, the electron-vibration interaction (EVI) \cite{PRB_Vibronic_paper} may occur and play an important role in the spin-dependent transport through dsDNA.

In this work, we focus on the influence of the EVI on the CISS effect and the spin-polarized transport in a dsDNA molecule, which hosts the helical chain-induced SOC, the environment-induced {dephasing process, the interchain and intrachain hopping integrals and the onsite vibration modes}, by using the Landauer-B{\"u}ttiker formula \cite{PRB_paper_Sun_1,PRB_paper_Sun_2,PRB_Fu_1}. Our theoretical investigations uncover that the EVI may enhance the spin polarization in the dsDNA molecule, and also bring a series of new spin-splitting transmission modes in the transmission spectra. {In some special structures, the vibration-induced additional resonance tunnelings lead to the spin-dependent transport in the band-gap regime of dsDNA molecules. Interestingly, these new spin-dependent transmission modes possess the same spin polarization as that in the original transmission modes. Moreover, the vibration-induced spin-dependent transport behaviors and the related spin polarization are rather robust against the the increasing dephasing. Although these theoretical results are obtained at zero temperature, they provide the fundamental understanding of the influences of EVI on the CISS effect and the spin-dependent transport in dsDNA molecules, and suggest a feasible route to detect the vibration-induced spin polarization in low-dimensional helical molecules.}

The remainder of this paper is organized as follows. In Sec. \Rmnum{2}, we construct a dsNDA-based spintronic device, {with a Hamiltonian model to describe the EVI, SOC, electronic hopping and dephasing process, and then introduce theoretical methods to study the spin-dependent transport}. In Sec. \Rmnum{3}, the spin-dependent transmission spectra and the related spin polarization in the dsDNA molecule are calculated, and the influences of the EVI on the CISS effect and spin-dependent transport are discussed in details. Finally, the main results are summarized in the last section.

\begin{figure}
\centering
\includegraphics[width=8.5cm]{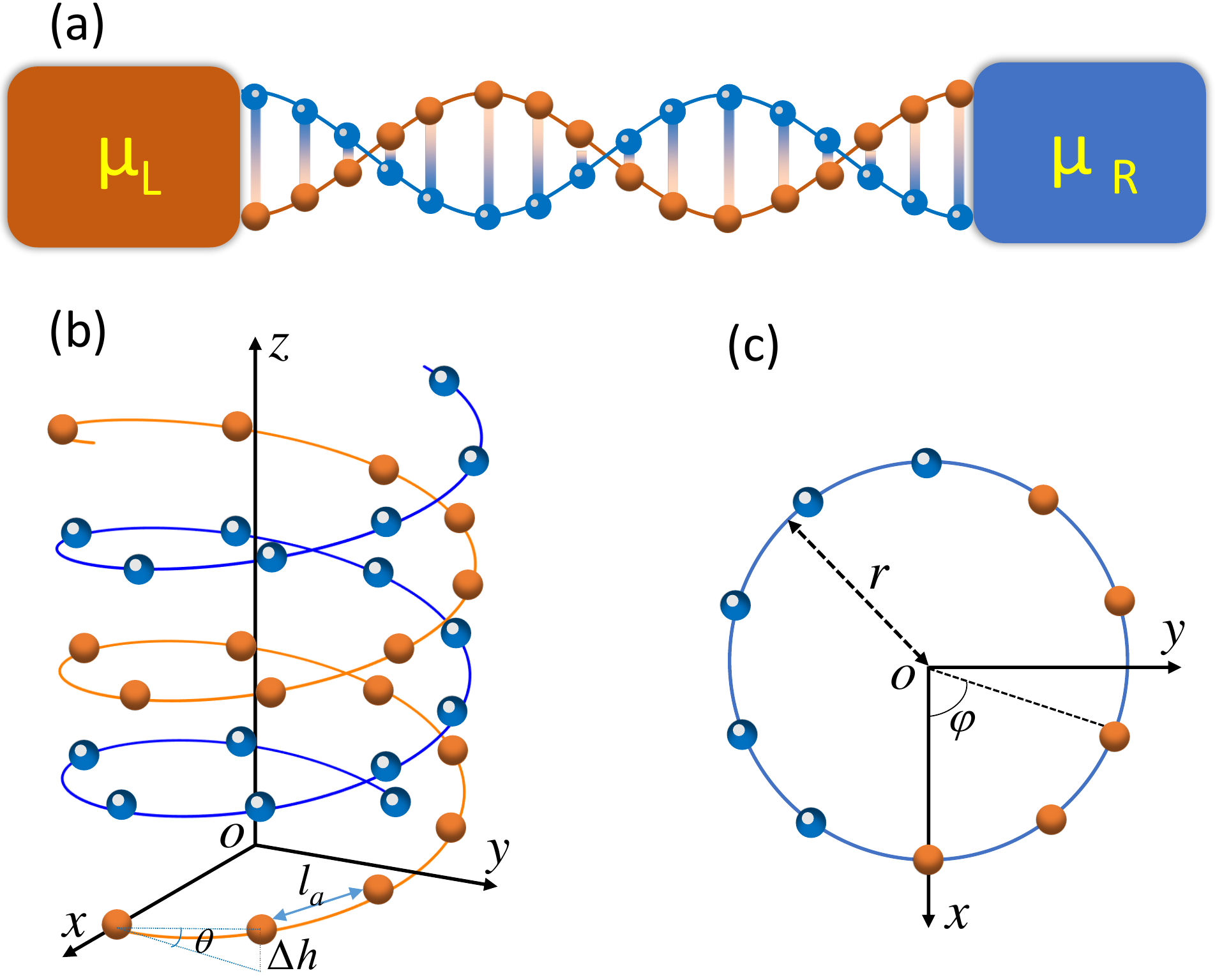}
\caption{(a) The schematic illustration of a chiral dsDNA molecule connected with two nonmagnetic leads. (b) Schematic view of right-banded dsDNA molecule along the $z$ direction. Here, the structural parameters $h$, $\theta$ and $l_a$ denote the pitch, helix angle and the length, respectively. (c) Projection of bottom five base pairs of the dsDNA molecule into the \emph{x}-\emph{y} plane with the radius $r$ and the twist angle $\Delta\varphi$.}
\label{fig1}
\end{figure}

\section{\Rmnum{2}. MODEL AND METHODS}

\subsection{A. Hamiltonian model of dsDNA-based device}
We construct a chiral chain-based dsDNA molecule coupled with two nonmagnetic leads, as illustrated schematically in Fig.~\ref{fig1}(a). In the central dsDNA molecule, we consider the interchain and intrachain {hopping integrals}, the symmetry-induced SOC, the EVI, the onsite vibration modes and the environmental-induced dephasing in the spin-selective transport process. The model Hamiltonian of this dsDNA-based device can be described as \cite{PRL_paper_Sun_1,PRB_paper_Sun_1}
\begin{equation}\label{H}
{\mathcal{H}=\mathcal{H}_{DNA}+\mathcal{H}_{el}+\mathcal{H}_{e-v}+\mathcal{H}_d+\mathcal{H}_{ph}},
\end{equation}
{where $\mathcal{H}_{DNA}$ is the Hamiltonian of usual two-leg model including the spin degree of freedom \cite{PRL_paper_Sun_1}, and can be described as $\mathcal{H}_{DNA}$ = $\mathcal{H}_{nt}+\mathcal{H}_{so}$. Here, $\mathcal{H}_{nt}$ = $\hat{\textbf{p}}^2/2m_e+V$ describes the kinetic and potential energies of electrons of the molecule, and $\mathcal{H}_{so}$ = ($\hbar/4m_e^2c^2$)$\nabla V\cdot$($\hat{\sigma}\times \hat{\textbf{p}}$) is the Hamiltonian of the SOC term; $\hbar$ is the reduced Plank constant, $c$ is the speed of light, $\hat{\textbf{p}}$ is the momentum operator, and $\hat{\sigma}$ = ($\sigma_x$, $\sigma_y$, $\sigma_z$) are the Pauli matrices. By using the second quantization as described in Ref. \cite{PNAS_Sun_paper}, $\mathcal{H}_{DNA}$ can be written as}
{
\begin{equation}\label{H_{DNA}}
\begin{split}
\mathcal{H}_{DNA}=&\sum\limits_{m,n} \Big[\varepsilon_{mn}c_{mn}^\dag c_{mn}+i\gamma_{mn}c_{mn}^\dag(\sigma_m^{n}+\sigma_m^{n+1})c_{m,n+1} \\
&+c_{mn}^\dag [t_{mn}+i\gamma_{mn}(\sigma_{m,z}^{n}+\sigma_{m,z}^{n+1})]c_{m,n+1}\\
&+\lambda c_{1n}^\dag c_{2n}+\mathrm{H}.\mathrm{c}.\Big].
\end{split}
\end{equation}} \\
Here, {$c_{mn}^\dag$ = ($c_{mn\uparrow}^\dag$, $c_{mn\downarrow}^\dag$) and $\varepsilon_{mn}$ are the creation operator and the onsite energy at the lattice site $\{m,n\}$ in the dsDNA molecule, denoting the site $n$ in chain $m$ (= 1 or 2)}. $t_{mn}$ and $\lambda$ are the intrachain and interchain hopping integrals. $\gamma_{mn}$ is the SOC parameter with the expression $\gamma_{mn}$ = $-\frac{\alpha}{4l_a}$, {where $\alpha\equiv\frac{\hbar^2}{4m^2c^2}\langle\frac{d}{dr}V(r)\rangle$. As $V(r)$ varies most rapidly in the near nuclear region, it is reasonable to consider its radical component only \cite{PRL_paper_Sun_1}. $\sigma_{m=1}^{n+1}$ = ($\sigma_x\sin\varphi\sin\theta-\sigma_y\cos\varphi\sin\theta$)($n\Delta\varphi$) and $\sigma_{m=2}^{n+1}$ = ($\sigma_x\sin\varphi\sin\theta-\sigma_y\cos\varphi\sin\theta$)($n\Delta\varphi+\pi$). $\sigma_{m=1,z}^{n+1}$ = $\sigma_z\cos\theta$($n\Delta\varphi$) and $\sigma_{m=2,z}^{n+1}$ =$\sigma_z\cos\theta$($n\Delta\varphi+\pi$)}. In the above expressions, $l_a$ and $\Delta\varphi$ are the arc length and the twist angle between successive base-pairs respectively, as described in Figs.~\ref{fig1}(b) and~\ref{fig1}(c).

{The Hamiltonian $\mathcal{H}_{el}=\mathcal{H}_{lead}+\mathcal{H}_{T}$, where $\mathcal{H}_{lead}=\sum_{k}\varepsilon_{k}(a_{\mathrm{L}k}^\dag a_{\mathrm{L}k}+a_{\mathrm{R}k}^\dag a_{\mathrm{R}k})$ describes the electrons in the left and right nonmagnetic leads, and $\mathcal{H}_T=\sum_{k}[\tau_{\mathrm{L}}a_{\mathrm{L}k}^\dag (c_{11}+c_{21})+\tau_{\mathrm{R}}a_{\mathrm{R}k}^\dag (c_{1N}+c_{2N})+\mathrm{H}.\mathrm{c}.]$ describes the electronic coupling between the dsDNA and the both leads with the strength $\tau_{\mathrm{L}}=\tau_{\mathrm{R}}=\tau$, where $a_{\mathrm{L}(\mathrm{R})k}^\dag$ is the creation operator for an electron in the left (right) lead, and $N$ represents the length of the dsDNA chain.}


The EVI Hamiltonian $\mathcal{H}_{e-v}$ can be described as \cite{JCP_evi}
{\begin{equation}\label{H_e-v}
\begin{split}
\mathcal{H}_{e-v}=\sum\limits_{mm'}\left[\sum\limits_{nn'}\Lambda_{m'n'mn} c_{m'n'}^\dag c_{mn}(a^\dag+a)\right].
\end{split}
\end{equation}} \\
Here, $a^\dag$ ($a$) is the creation (annihilation) operator for the phonon mode. {In the following analysis, we assume that the coupling parameters are set to be nonzero values only in the three cases: (i) for the onsite electrons in the dsDNA chain (\emph{i.e.}, $m'$ = $m$, $n'$ = $n$), $\Lambda_{m'n'mn}$ is reduced to $\Lambda$; (ii) for the intrachain nearest-hopping electrons (\emph{i.e.}, $n'$ = $n\pm1$, $m'$ = $m$), $\Lambda_{m'n'mn}$ is set as $M_1$ and (iii), for the interchain nearest-hopping electrons (\emph{i.e.}, $n'$ = $n$, $m$ = 1, $m'$ = 2), $\Lambda_{m'n'mn}$ is set as $M_2$. It is reasonable to postulate that $M_{1(2)}<\Lambda$ in the dsDNA molecules. In our calculations, $M_1=M_2=0.2\Lambda$ is adopted.}


The forth term in Eq. (1), $\mathcal{H}_d$, is the Hamiltonian of the B\"{u}ttiker virtual leads and its coupling with each base of the dsDNA \cite{PRB_Buttiker_1,PRB_Buttiker_2,JCP_Kilgour}, {simulating the phase-breaking processes due to the inelastic scattering with phonons and counterions \cite{JACS_Butter_lead,CP_Butter_lead}}. In the frame of the tight-binding model, $\mathcal{H}_d$ is expressed as
{\begin{equation}\label{H_d}
\mathcal{H}_d=\sum_{mnk}(\varepsilon_{mnk}d_{mnk}^\dag d_{mnk}+t_dd_{mnk}^\dag c_{mn}+\mathrm{H}.\mathrm{c}.),
\end{equation}}\\
where $d_{mnk}^\dag$ = ($d_{mnk\uparrow}^\dag$, $d_{mnk\downarrow}^\dag$) and $\varepsilon_{mnk}$ describe the creation operator and onsite energy of mode $k$ in B\"{u}ttiker virtual leads, {and $t_d$ is the coupling between the nucleobase and the virtual lead.}

The last term, $\mathcal{H}_{ph}=\omega_0{a^\dag}a$, represents the vibrational mode with the phonon frequency $\omega_0$.

\subsection{B. Lang and Firsov transformation}
It is noted that we may eliminate the EVI term from the Hamiltonian (1) by employing the commonly used small polaron (Lang and Firsov) transformation \cite{PRB_polaron_1,PRB_polaron_2,PRB_polaron_3}, which converts the Hamiltonian $\mathcal{H}$ into the form $\tilde{\mathcal{H}}$ = $e^S \mathcal{H} e^{-S}$ with $S$ = $\frac{\Lambda}{\omega_0}\sum\limits_{m,n}c_{mn}^\dag c_{mn}$($a^\dag-a$)+$\sum\limits_{m,n}$($\frac{M_1}{\omega_0}c_{mn}^\dag c_{m,n+1}+\frac{M_2}{\omega_0}c_{1n}^\dag c_{2n}$)($a^\dag-a$). The transformed Hamiltonian reads as {$\tilde{\mathcal{H}}$ = $\tilde{\mathcal{H}}_{DNA}+\tilde{\mathcal{H}}_{el}+\tilde{\mathcal{H}}_d+\tilde{\mathcal{H}}_{ph}$}, where the vibration term remains unchanged, while the electron part $\mathcal{H}_{DNA}$ is reshaped into
{
\begin{equation}\label{H_{DNA}}
\begin{split}
\tilde{\mathcal{H}}_{DNA}=& \sum\limits_m \Big[\sum\limits_{n=1}^N (\tilde{\varepsilon}_{mn}c_{mn}^\dag c_{mn}+\tilde{\lambda} c_{1n}^\dag c_{2n})  \\
&+\sum\limits_{n=1}^{N-1}\big[i\tilde{\gamma}_{mn} c_{mn}^\dag(\sigma_m^{n}+\sigma_m^{(n+1)})c_{m,n+1}  \\
&+c_{mn}^\dag(\tilde{t}_{mn}+i\tilde{\gamma}_{mn}(\sigma_{m,z}^{n}+\sigma_{m,z}^{(n+1)})) c_{m,n+1} \big] \\
&+\mathrm{H}.\mathrm{c}.\Big].
\end{split}
\end{equation}
}

It is clear that due to the EVI, the energy level of the dsDNA molecule is renormalized to $\tilde{\varepsilon}_{mn}\equiv\varepsilon_{mn}-\frac{\Lambda^2}{\omega_0}$. Assuming that EVI is sufficiently weak, \emph{i.e.}, $\Lambda\ll\gamma_{mn}$, the coupling parameter {$\gamma_{mn}$} can be renormalized in a similar way: {$\tilde{\gamma}_{mn}$ = $\gamma_{mn}-\frac{2\Lambda M_1}{\omega_0}$}. Similarly, we can obtain $\tilde{t}_{mn}$ = $t_{mn}-\frac{2\Lambda M_1}{\omega_0}$, $\tilde{\lambda}$ = $\lambda-\frac{\Lambda M_2}{\omega_0}$. The dressed tunneling matrix elements are transformed to {$\widetilde{\tau}_{\mathrm{L}(\mathrm{R})}\equiv \tau_{\mathrm{L}(\mathrm{R})} X$, $\widetilde{t}_d\equiv t_d X$. Then, the Hamiltonian $\mathcal{H}_{el}$ is transformed to $\widetilde{H}_{el}=\sum_{k}[\widetilde{\varepsilon}_{k}(a_{\mathrm{L}k}^\dag a_{\mathrm{L}k}+a_{\mathrm{R}k}^\dag a_{\mathrm{R}k})+ \widetilde{\tau}_{\mathrm{L}}a_{\mathrm{L}k}^\dag (c_{11}+c_{21})+\widetilde{\tau}_{\mathrm{R}}a_{\mathrm{R}k}^\dag (c_{1N}+c_{2N})+\mathrm{H}.\mathrm{c}.]$, and $\mathcal{H}_d$ is converted to be $\widetilde{\mathcal{H}}_d$ = $\sum_{mnk}(\widetilde{\varepsilon}_{mnk}d_{mnk}^\dag d_{mnk}+\widetilde{t}_dd_{mnk}^\dag c_{mn}+\mathrm{H}.\mathrm{c}.)$, where $\widetilde{\varepsilon}_{k}=\varepsilon_{k}-\frac{\Lambda^2}{\omega_0}$ and $\widetilde{\varepsilon}_{mnk}=\varepsilon_{mnk}-\frac{\Lambda^2}{\omega_0}$}. Note that the phonon operator $X\equiv {\rm exp}\left[-\left(\frac{\Lambda}{\omega_0}\right)(a^\dag-a)\right]$ arises from the canonical transformation of the particle operator $e^S{\rm \emph{d}} e^{-S}$ = ${\rm \emph{d}}X$ \cite{PRB_polaron_1,Theo_1}. For this model, we assume that the vibrational mode is coupled with a thermal phonon bath and that this coupling is strong enough so that the phonon maintains its thermal equilibrium state throughout the process. Therefore, the expected value of the phonon operator $X$ can be expressed as the following one \cite{PhysRevB.71.165324,PRB_Imry}
\begin{equation}\label{X}
\langle X\rangle={\rm exp}\left[-\left(\frac{\Lambda}{\omega_0}\right)^2(N_{ph}+\frac{1}{2})\right],
\end{equation}
where $N_{ph}$ denotes the equilibrium phonon population. Here, we consider the low temperature regime where $kT\ll\Lambda$, $\omega_0$, so $\langle X\rangle$ can be approximated by ${\rm exp}[-\frac{1}{2}(\frac{\Lambda}{\omega_0})^2]$, which is independent of temperature. Thus, we can decouple the electron and phonon subsystems by replacing $X$ with its expectation value $\langle X\rangle$.


\subsection{C. Spin-dependent transmission calculations}
Using the Landauer-B$\ddot{\mathrm{u}}$ttiker formula \cite{Ryndyk2009,ryndyk2016theory,NEFG_1,NEFG_2}, {the spin-dependent transmission coefficient of the dsDNA molecule from the $p$th lead with spin $s'$ to the $q$th with spin $s$ can be calculated as
\begin{equation}\label{T}
\begin{aligned}
T_{qs,ps'}=\mathrm{Tr}[\Gamma_{qs}{G^R_{qs,ps'}}(\varepsilon)\Gamma_{ps'}(\varepsilon)G^A_{qs,ps'}(\varepsilon)],
\end{aligned}
\end{equation}
where $G^{R(A)}_{qs,ps'}(\varepsilon)$ are the retarded (advanced) Green's functions for spin-up or spin-down electrons \cite{mahan2013many}. $\Gamma_{qs}$ and $\Gamma_{ps'}$ are the linewidth functions describing the coupling between the leads and the dsDNA molecule, where $\varepsilon$ is the incident electron energy (Fermi energy), and $\Gamma_{qs}=i[\Sigma^r_{qs}-\Sigma^a_{qs}]$ with $\Sigma^{r(a)}_{qs}$ the retarded (advanced) self-energy due to the coupling to the \emph{q}th lead. For the real left/right lead, $\Sigma^r_{(\mathrm{L}/\mathrm{R})s}=-i\Gamma_{\mathrm{L}/\mathrm{R}}/2=-i\pi\rho_{\mathrm{L}/\mathrm{R}}\widetilde{\tau}_{\mathrm{L}/\mathrm{R}}^2$; while for the virtual leads, $\Sigma^r_{qs}=-i\Gamma_d/2=-i\pi\rho_d\widetilde{t}_d^2$, with the dephasing parameter $\Gamma_d$ and $\rho_{\mathrm{L}/\mathrm{R}/d}$ being the density of states of the leads. Note that as we calculate the spin-up (-down) transmission spectra from the real left lead ($p=1$) to the real right one ($q=N$), the spin-up (-down) transmission coefficient is simplified as $T_{up(dn)}$ for convenience}. Thus, the spin polarization in the dsDNA molecule is defined as $P_s$ = $\left(T_{up}-T_{dn}\right)/\left(T_{up}+T_{dn}\right)$.

\begin{figure*}[!t]
\centering
\includegraphics[width=18.0cm]{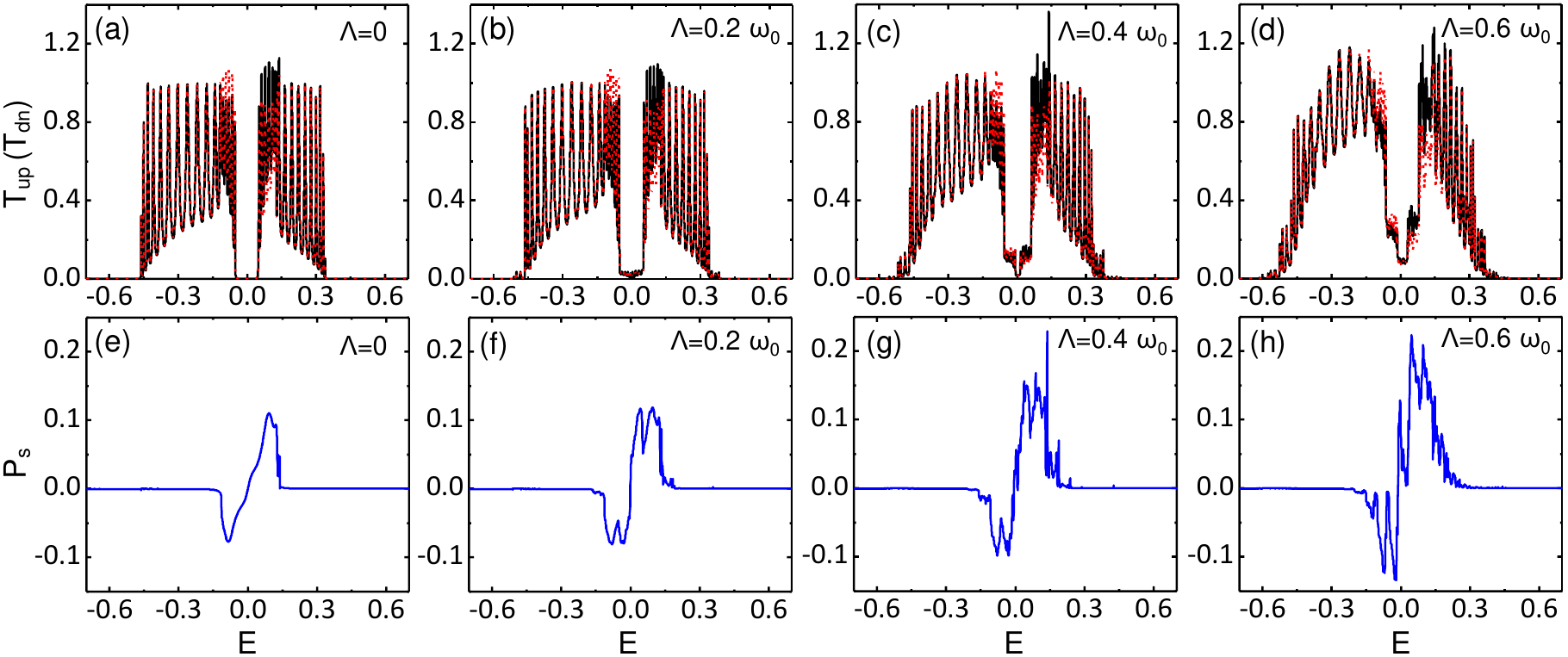}
\caption{(a)-(d) The spin-dependent transmission spectra $T_{up}$ and $T_{dn}$ versus the energy $E$ in the absence of EVI and without dephasing, while with the different values of $\Lambda$, which is set as 0, 0.2$\omega_0$, 0.4$\omega_0$ and 0.6$\omega_0$, respectively. The black solid line represents the spin-up transmission spectra, and the red dash line represents the spin-down ones. (e)-(h) The corresponding spin polarization $P_s$ versus $E$. The other structural parameters are set as $\omega_0$ = 0.05 eV, $\Gamma_d$ = 0 and $\gamma_{1n}$ = 0.01 eV.}
\label{fig2}
\end{figure*}

As interpreted above, when the operator $X$ is replaced by $\langle X\rangle$, the Hamiltonian can be decoupled from the vibration operator. {The electronic Green's functions on the Keldysh contour may be approximated as a product of the pure electronic term that can be computed based on the transformed Hamiltonian $\mathcal{\tilde{H}}$ and the Franck-Condon factor \cite{Franck_Condon_1,Franck_Condon_2,Franck_Condon_3},
\begin{equation}\label{Glesser}
\begin{aligned}
G^R_{qs,ps'}(t,t')&\approx{-\frac{i}{\hbar}}\langle{{T_c}c_{qs}(t) c^{\dag}_{ps'}(t')}\rangle_{\tilde{\mathcal{H}}}\langle{X(t)X(t')}\rangle \\
&=\tilde{G}^{R}_{qs,ps'}(t,t')e^{-\Phi(t-t')}.
\end{aligned}
\end{equation}
Similarly,
\begin{equation}\label{Gbiger}
G^A_{qs,ps'}(t,t')=\tilde{G}^A_{qs,ps'}(t,t')e^{-\Phi(t-t')},
\end{equation} \\
where the identification $e^{-\Phi(t-t')}$ = $\Sigma_{\eta=-\infty}^\infty L_{\eta} e^{-i\eta\varepsilon (t-t')}$. Here, the index $\eta$ represents the number of vibration phonons involved and $L_{\eta}$ are the coefficients, depending on the temperature and the strength of the EVI. At a finite temperature, $L_{\eta}$ can be expressed as
\begin{equation}\label{Ln1}
L_{\eta}=e^{-g\left(2N_{ph}+1\right)}e^{\eta\omega_0\beta/2}I_{\eta}\left(2g\sqrt{N_{ph}\left(N_{ph}+1\right)}\right),
\end{equation}} \\
where $g$ = $\left(\frac{\Lambda}{\omega_0}\right)^2$, $\beta=1/k_BT$, and $I_{\eta}$ is the modified Bessel function of the $\eta$th order. At the zero temperature, $L_{\eta}$ can be simply read
\begin{equation}\label{Ln2}
L_{\eta}\approx{\rm exp}\left[-\left(\frac{\Lambda}{\omega_0}\right)^2\right]\left(\frac{\Lambda}{\omega_0}\right)^{2|\eta|}\frac{1}{|\eta|!}.
\end{equation}

For clarity, the electronic parameters are considered uniform along each helix of the dsDNA molecule. Based on the complementary base-pairing rule, the dsDNA molecule consists of four nucleobases, \emph{i.e.}, guanine (G), adenine (A), cytosine (C), and thymine (T). Because the structure and the atom number of these nuclear bases are different, the electronic parameters between the two DNA strands may be asymmetrical \cite{doi:10.1063/1.1352035,doi:10.1021/ja054257e,Hawke2010}. {For the dsDNA molecule considered here, $\varepsilon_{mn}$ is set to $\varepsilon_{1n}=0$ and $\varepsilon_{2n}=0.3$, $t_{mn}$ is taken as $t_{2n}=0.1$ and $\lambda$ = -0.08. To describe the asymmetry between the two helical chains, we employ an additional parameter $x$, and set $t_{1n}$ = $xt_{2n}$, $\gamma_{2n}$ = $x\gamma_{1n}$ with $x$ = 1.4. All these parameters are extracted from first-principles calculations \cite{RevModPhys_1,doi:10.1063/1.1352035,doi:10.1021/ja054257e,Hawke2010} and the unit is eV. The SOC is estimated to $\gamma_{1n}$ = 0.01 eV, which is an order of magnitude smaller than the intrachain hopping integral. For the real leads, the parameters $\tau_{\text{L}}=\tau_{\text{R}}=1$ are fixed.} The remaining parameters are taken as $N$ = 20, $\theta$ = 0.66 rad, and $\Delta\phi$ = $\pi/5$, resembling the \emph{B}-form dsDNA molecule, in which the helix makes a turn every 3.4 nm, and the distance between two neighboring base pairs is 0.34 nm \cite{B_form_DNA}. Note that these parameters are used throughout this work, unless other values are explicitly mentioned.

\begin{figure*}
\centering
\includegraphics[width=18.0cm]{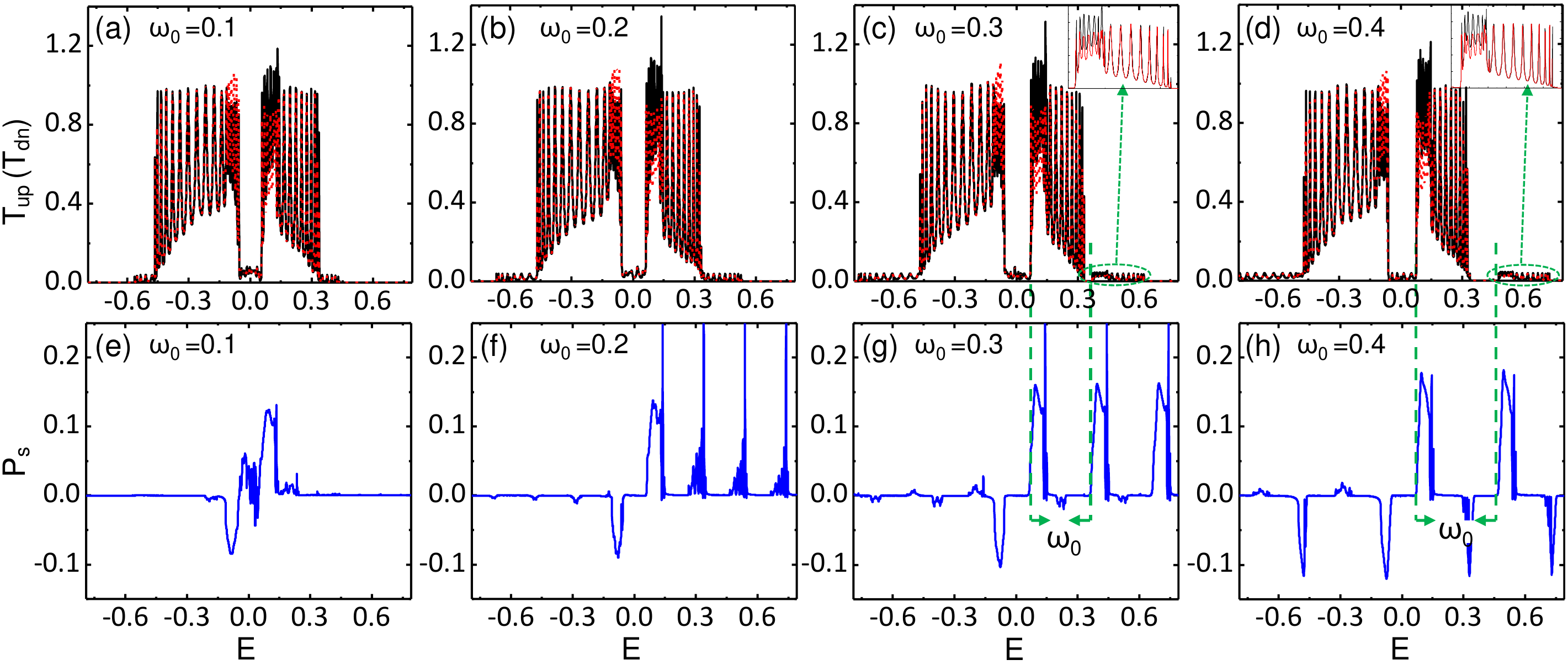}
\caption{(a)-(d) The spin-dependent transmission spectra $T_{up}$ and $T_{dn}$ versus the energy $E$ under the different values of vibration frequency $\omega_0$, where $\omega_0$ is set as 0.1, 0.2, 0.3 and 0.4 eV, respectively. The black solid line represents the spin-up transmission spectra and the red dash line represents the spin-down ones. (e)-(h) The corresponding $P_s$ of the above four cases. The other parameters are set as $\Lambda$ = 0.2 eV, $\Gamma_d$ = 0 and $\gamma_{1n}$ = 0.01 eV.}
\label{fig3}
\end{figure*}

\section{\Rmnum{3}. RESULTS AND DISCUSSION}

To illustrate the validity of our device model and theoretical method, we firstly calculated the spin-dependent transmission spectra of the dsDNA-based device without the EVI and dephasing process. The numerical results are drawn in Fig.~\ref{fig2}(a), where a weak SOC is adopted as $\gamma_{1n}=0.01$ eV, and the other parameters are set as $\Gamma_d$ = 0 and $\omega_0$ = 0.05 eV. One may see that the conductance spectrum is consisted of two transmission bands, \emph{i.e.}, the highest occupied molecular orbital (HOMO) and the lowest unoccupied molecular orbital (LUMO), which are divided by an energy gap. {For convenience, we call this energy gap the HOMO-LUMO gap}. Many transmission peaks are found in both HOMO and LUMO bands, and their number is strictly equal to that of the base-pairs in dsDNA chain due to the quantum coherence effect. {Meanwhile, the spin polarization $P_s$ occurs in the both HOMO and LUMO bands, especially near the edges of the HOMO-LUMO gap as drawn in Fig.~\ref{fig2}(e).} These properties are consistent with previous results \cite{PRB_paper_Sun_1}. {As the EVI is taken into account in the dsDNA molecule in a range 0.2$\omega_0\leq\Lambda\leq0.6\omega_0$, the CISS effect and the corresponding spin-dependent transport behavior changes remarkably as illustrated in Figs.~\ref{fig2}(b)-\ref{fig2}(d)}. In particular, one can identify several interesting EVI-induced spin-resolved transport features as listed in the following:

\begin{figure*}
\centering
\includegraphics[width=18.0cm]{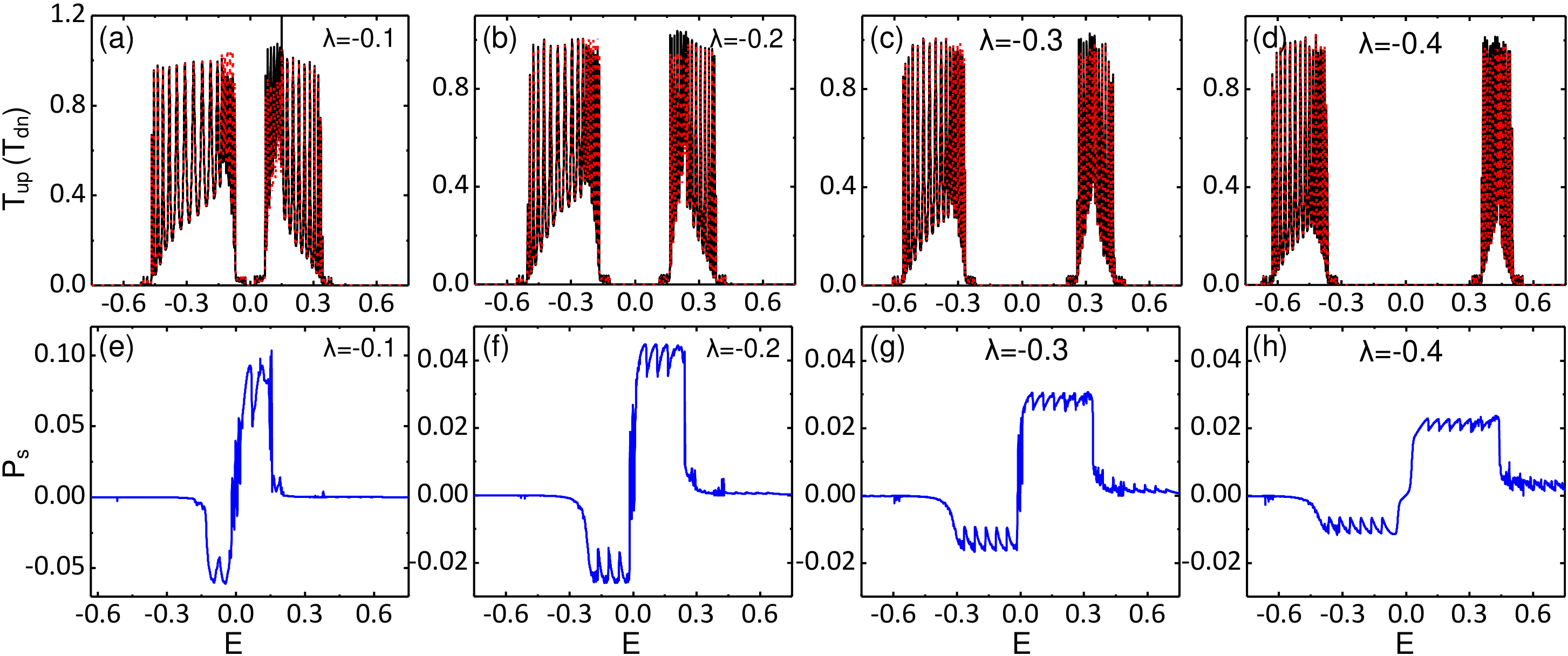}
\caption{(a)-(d) $T_{up}$ and $T_{dn}$ under the different value of $\lambda$. where the value is set as -0.1, -0.2, -0.3 and -0.4 eV, respectively. The black solid line represents the spin-up transmission spectra, the red dash line represents the spin-down transmission spectra. (e)-(h) The corresponding spin polarization $P_s$ versus the energy $E$. Where $\Lambda$ = 0.2$\omega_0$, $\Gamma_d$ = 0, $\omega_0$ = 0.05 eV, $\gamma_{1n}$=0.01 eV.}
\label{fig4}
\end{figure*}

(i) Some additional small transmission peaks appear in the central energy gap, {indicating that the EVI assists the electronic conduction through the dsDNA in its HOMO-LUMO gap}, while both spin-selective effect and electron-hole-type symmetry are hold in the transmission spectra. One may refer to Fig. S1 in Supplementary Material (SM) for more details \cite{Suppl_Mater}.

(ii) The transmission peaks near the Fermi level in both HOMO and LUMO bands are enhanced as $\Lambda$ increases, {indicating that the EVI contributes to charge transport through the dsDNA molecule, due to the vibration-induced additional conductance tunnelings}.


(iii) The spin polarization $P_s$ is also enhanced with the increasing of the EVI as shown in Figs.~\ref{fig2}(e)-\ref{fig2}(h), supporting that EVI strengthens the CISS effect in the dsNDA molecule. {This is because that the EVI enhances the electronic coherent effect, bringing more tunneling paths in the dsDNA molecule}. This indicates that we may attain a new physical mechanism to enhance the CISS effect and the spin-dependent transport in dsDNA and related molecules.

To elucidate the vibration-mediated spin-selective transport in the dsDNA molecule, we study the influence of the vibration frequency on the CISS effect in the dsDNA molecule. Figs.~\ref{fig3}(a)-\ref{fig3}(h) plot the spin-dependent transmission spectra $T_{up}$ and $T_{dn}$ versus the Fermi level $E$ with increasing $\omega_0$ in the absence of the dephasing ($\Gamma_d$ = 0). Similarly, one may find that some new resonance peaks in the HOMO-LUMO gap, driving the dsDNA molecule to a conductor. Nevertheless, their heights are much lesser than the peaks in the main transmission spectra. Moreover, both outsides of the HOMO and LUMO bands show additional small transmission peaks, remaining the spin-splitting characteristics regardless the change of vibration frequencies. It should be stressed that the new transmission modes in the high-energy regime nearly duplicate the transmission features of the LUMO band, as highlighted in the insets of Figs.~\ref{fig3}(c) and~\ref{fig3}(d). In comparison with the energy $\tilde{\varepsilon}$ of the main transmission mode, the vibration-induced transmission mode is located just at $E=\tilde{\varepsilon}+\omega_0$, confirming that the new transmission modes originate from the vibration resonance. Consequently, as $\omega_0$ increases, the new transmission modes shift towards the high-energy direction while maintaining the all spin transport characteristics, as shown in Fig.~\ref{fig3}(d). For the convenience in the following discussions, we call the main resonance mode for $E > 0$ as the 0$^{th}$ transmission, and the new transmission mode at $E=\tilde{\varepsilon}+\omega_0$ as the 1$^{st}$ transmission. {Note that every peak in the 1$^{st}$ transmission mode serves as an extra conductance channel for electrons in the dsDNA molecule}. From the general feature of the EVI, the vibration may induce a series of higher-order transmission modes, and the $n^{th}$ tunnelling peak in the $m^{th}$ modes should be localized at $E_{mn}=\tilde{\varepsilon}+(m-1+\frac{n}{N})\omega_0$, with $m=0,1,...$ and $n=1,...,N$. Nevertheless, due to tiny heights in these resonance peaks, it is difficult to observe them in the transmission spectra. In addition, it is noted that for $E<0$, the vibration-induced 1$^{st}$ and higher-order transmission modes appear at $E_{mn}=\tilde{\varepsilon}-(m-1+\frac{n}{N})\omega_0$, where $\tilde{\varepsilon}$ denotes the energy related to the 0$^{th}$ transmission mode in the HOMO band.

Now we turn to examine the influence of vibration frequency $\omega_0$ on the spin polarization $P_s$ in the dsDNA molecule. In Figs.~\ref{fig3}(e)-\ref{fig3}(h), the corresponding $P_s$ versus $E$ with increasing $\omega_0$ are drawn. One may find that the $P_s$ spectra display several interesting characteristic properties: (i) As $\omega_0$ increases, $P_s$ mode related to the 0$^{th}$ transmission spectrum is enhanced, indicating that the vibration indeed enhances the CISS effect in chiral molecules. {In fact, the EVI proposed here does not alter the SOC, nor break helical symmetry and the spin memory in dsDNA molecules, thus the origination of the spin splitting maintains well in the system}. Moreover, the vibration brings several new resonance tunneling paths, much likes the situation that the increasing length of dsDNA chain produces more transmission peaks, which have already been confirmed by previous theoretical calculations and experimental observations \cite{Science_paper,PRL_paper_Sun_1}. Consequently, $P_s$ is enhanced remarkably by the increase of vibration frequency in the present device models. (ii) With increasing $\omega_0$, a series of new $P_s$ modes, such as the 1$^{st}$, 2$^{nd}$, and even 3$^{rd}$ mode, appear around $E=\tilde{\varepsilon}\pm{n\omega_0}$, and meanwhile, the heights of \emph{n}$^{th}$ ($n\geq{1}$) $P_s$ mode is enhanced to catch up to that of the 0$^{th}$ $P_s$ mode. As a result, all $P_s$ modes host the same shapes including their heights and widths, as illustrated in Figs.~\ref{fig3}(g) and \ref{fig3}(h). This can be considered as a new finding in the field of chiro-spintronics, since it has not been reported so far. (iii) As $\omega_0$ increases to a large value, such as $\omega_0$ = 0.4 eV, some sharp valleys with $P_s < 0$ appear and approach to the one in the 0$^{th}$ $P_s$ mode. This is due to the fact at every vibration-induced transmission spectrum reproduces the one in both HOMO and LUMO bands, and the resonance valleys are associated with the spin polarization in the HOMO band. For small values of $\omega_0$, the vibration-induced transmission peaks related to these $P_s$ valleys are buried under the transmission spectra in the LUMO band.

\begin{figure}
\centering
\includegraphics[width=8.6cm]{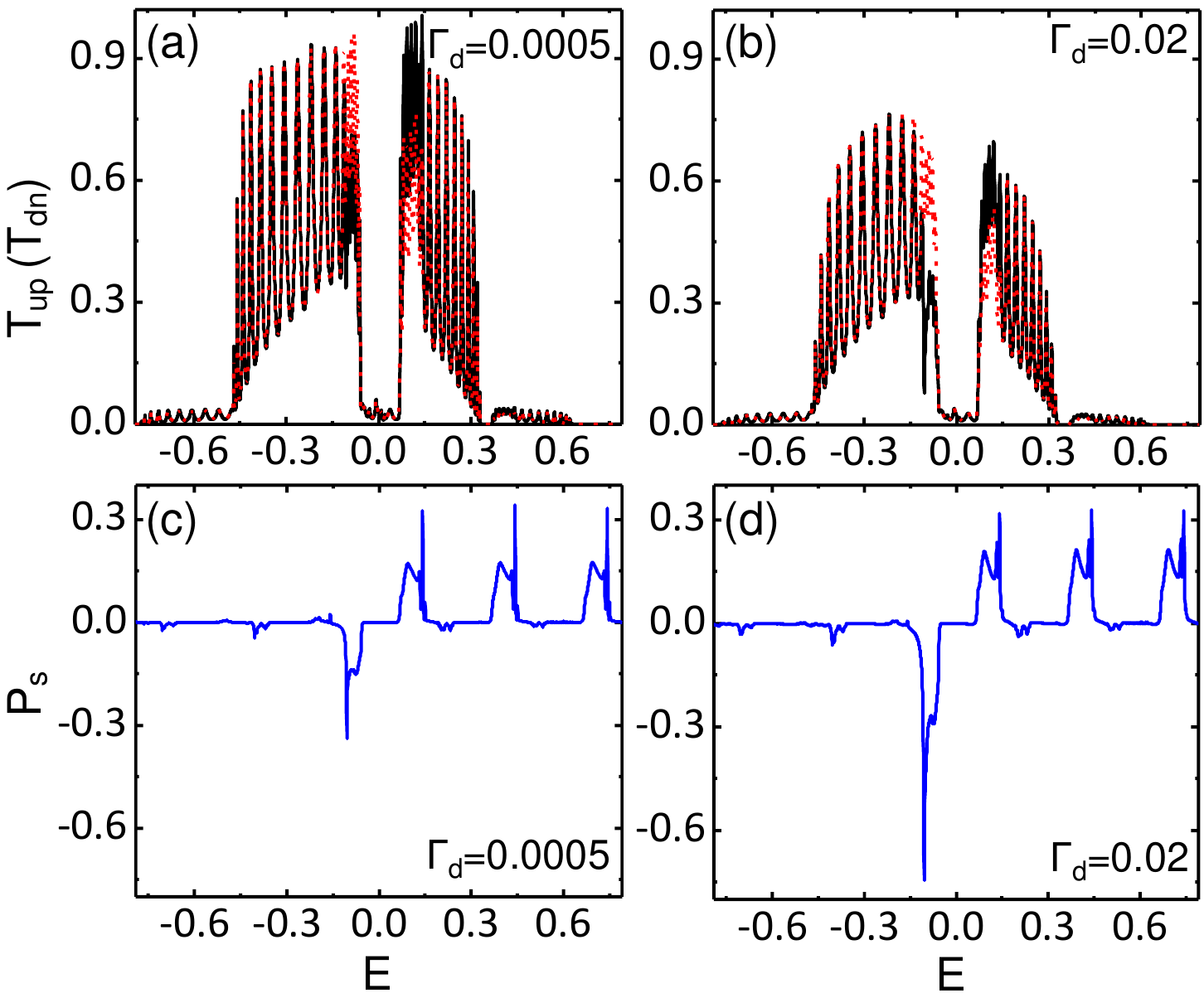}
\caption{(a) and (b) $T_{up}$ and $T_{dn}$ under the several values of $\Gamma_d$, which is set as 0.0005 and 0.02 eV, respectively. The black solid line represents the spin-up transmission spectra, the red dash line represents the spin-down transmission spectra. (c) and (d) The corresponding spin polarization $P_s$ versus the energy $E$. The other structural parameters are set as $\Lambda=0.2\omega_0$, $\gamma_{1n}=0.01$ eV, $\lambda=-0.08$ eV and $\omega_0=0.3$ eV.}
\label{fig5}
\end{figure}

To illustrate the exotic finding that all vibration-induced spin-dependent transmission spectra possess the same $P_s$ modes, we consider further a particular dsDNA-based device model, in which the electronic hopping parameter $\lambda$ in the dsDNA molecule is changed from -0.1 to -0.4 eV, while $\omega_0$ and $\Lambda$ are fixed as $0.05$ eV and $0.2\omega_0$, respectively. The corresponding spin-dependent transmission spectra are plotted in Figs.~\ref{fig4}(a)-\ref{fig4}(d). It is obvious that the increase of hopping leads to a larger band gap between the HOMO and LUMO bands, and meanwhile, more narrow conductance plateaus emerge in the transmission spectra. For the dsDNA molecule, electron can transport not only along the helical chain, but also within the base-pairs. As $\lambda$ increases, the electronic localization in every base-pair is enhanced and charge transport along the main chains is frustrated. More interestingly, two asymmetrical platforms characterized by sawtooth shapes appear in the $P_s$ modes, as described in Figs.~\ref{fig4}(e)-\ref{fig4}(h). In particular, the appearance of the $P_s$ platform in the band gap indicates that there is filled with the vibration-induced spin-splitting tunneling peaks, although they are much difficult to observe in the transmission spectra. To support this derivation, in Fig. S2 in the SM \cite{Suppl_Mater}, we redrawn the spin-dependent transmission spectra in the band-gap regime for more details. A series of vibration-produced spin-dependent transmission spectra are indeed filled in the band gap, while their transmission values decrease nearly by two orders of magnitude between two neighboring spectra towards the zero-energy direction. However, every new transmission spectrum hosts the same $P_s$ values, confirming further the aforementioned conclusion. Owning to the fact the frequency is too small set in the present device, the vibration-induced transmission spectra are difficult to display in the bang gap. Moreover, all spin-dependent transmission spectra contact with each other with the same $P_s$ modes, making the $P_s$ spectra to display as a wide platform. {This property puts forwards a feasible route for us to detect the spin polarization produced by the vibrations in the chiral-molecule systems}. In addition, as interpreted above, the increasing $\lambda$ hinders the electrons to transport along the main helical chains, which reduces the SOC according to its Hamiltonian $H_{so}$. As a result, the $P_s$ platform decreases obviously with increasing $\lambda$.

It is well known that the dephasing process occurs inevitably in dsDNA molecule in experiments, thus it is natural to ask whether the dephasing breaks the vibration-induced CISS effect and decreases the spin polarization in the dsDNA molecule. In Figs.~\ref{fig5}(a) and \ref{fig5}(b), we plotted the spin-dependent transmission spectra versus $E$ for two values of the dephasing parameter $\Gamma_d$ in the presence of the EVI ($\Lambda=0.2\omega_0$) and factors $\gamma_{1n}=0.01$ eV, $\lambda=-0.08$ eV and $\omega_0=0.3$ eV. Note that as $\Gamma_d$ increases to large values (two additional cases for $\Gamma_d$ values are supplemented in Fig. S4 \cite{Suppl_Mater}), the spin-dependent transmission bands $T_{up}$ and $T_{dn}$ quickly decrease, because the dephasing process gives rise to the lose of electrons' phase and spin memory. As a result, the coherence of the dsDNA molecule is reduced and meanwhile, the oscillation peaks in the transmission spectra decrease remarkably. However, the 1$^{st}$ and even higher-order vibration-induced spin-dependent transmission modes still remain in the transmission spectra, indicating that the decoherence effect on the vibration-induced transmission modes is much less than that on the main transmission spectra. Moreover, the $P_s$ modes related with the vibration-induced transmission spectra are barely influenced by the increase of $\Gamma_d$, as illustrated in Figs.~\ref{fig5}(c) and \ref{fig5}(d), which suggests the robustness of the vibration-induced spin-splitting transmissions in the dsDNA molecule. Moreover, numerical results show that the dephasing process may even enhance the spin polarization in the HOMO bands. To support the conclusions obtained above, we also considered two other dsDNA-based spintronics devices in the presence of dephasing (see Figs. S5 and S6 in the SM \cite{Suppl_Mater}), the same spin-dependent transport properties are achieved and the robustness of the vibration-induced spin-splitting transmission spectra are confirmed further.

\section{\Rmnum{4}. CONCLUSION}

In summary, we investigate the influence of EVI on the CISS effect and the spin-dependent transport properties of the dsDNA molecule by considering helical symmetry-induced SOC, the dephasing process, and the interchain and intrachain hoppings with the Landauer-B{\"u}ttiker formula. Our theoretical results show that the EVI not only enhances the spin polarization but also produces a series of new spin-dependent transmission channels through the dsDNA molecule. The vibration-induced spin-transmission spectra tend to retain the same spin-polarization mode as in the main spin-transmission spectra, making the spin-polarization spectra to display a series of platforms even in the band gap. {Moreover, the vibration-induced spin-dependent transmissions are robust against the dephasing process, assisting the CISS and spin selective transport. These theoretical results provide new insights to understand the influence of the EVI on the CISS effect and the spin-polarized transport in dsDNA molecules. They also put forward a feasible route to enhance spin polarization induced by vibrations in low-dimensional molecular systems for applications}.

\section{ACKNOWLEDGNEBTS}
This work is supported by the Natural Science Foundation of China with Grants No. 11774104. Work at UCI was supported by DOE-BES (Grant No. DE-FG02-05ER46237). Computer simulations were partially performed at the U.S. Department of Energy Supercomputer Facility (NERSC).

\bibliography{REF}

\clearpage


\end{document}